\documentclass[11pt]{article}

\usepackage{amsmath}
\usepackage{amssymb}
\usepackage{latexsym}
\usepackage{graphics}
\usepackage{psfrag,fancyhdr,epsfig}
\usepackage{cite}
\newcommand{\mb}[1]{\mathbf{#1}}

\newcommand{\gap}{\,\,}

\newcommand{\Fu}{\mathcal{F}}
\newcommand{\Gu}{\mathcal{G}}

\newcommand{\Hu}{\mathcal{H}}

\newcommand{\Wu}{\mathcal{W}}

\newcommand{\Yu}{\mathcal{Y}}

\newcommand{\Phibar}{\overline{\Phi}}
\newcommand{\phibar}{\overline{\phi}}
\usepackage{color}

\definecolor{orange}{rgb}{0.9,0.2,0}
\definecolor{brown}{rgb}{0.7,0.3,0.2}
\definecolor{fuxia}{rgb}{1,0,1}
\definecolor{skyblue}{rgb}{0,0.1,0.9}
\definecolor{violetred}{rgb}{0.8,0.13,0.56}
\definecolor{deeppink}{rgb}{1.00,0.08,0.5}
\definecolor{pink}{rgb}{1.00,0.75,0.80}
\definecolor{orchid}{rgb}{0.85,0.44,0.84}
\definecolor{lightpink}{rgb}{1.00,0.71,0.76}
\definecolor{bluish}{rgb}{0,0.6,0.8}
\begin{document}

\begin{titlepage}

\begin{centering}

\hfill hep-ph/yymmnnn\\

\vspace{1 in}
{\large{\bf {Proton Stability in $\bf{SU(5)\times U(1)}$ and $\bf{SU(6)\times SU(2)}$ GUTs} }}\\
\vspace{1 cm}
{A.E. Faraggi${}^{1}$, M. Paraskevas${}^{2}$, J. Rizos${}^{2}$
 and K. Tamvakis${}^{2}$}\\
\vskip 0.5cm
$1\,$ Department of Mathematical Sciences, University of Liverpool\\
Liverpool L69 7ZL, UK\\
$2\,$ Physics Department, University of Ioannina\\
GR451 10 Ioannina, Greece

\vskip 0.5cm

\vspace{1.5cm}
{\bf Abstract}\\

\end{centering}
\vspace{.1in}

We consider explicit unified models based on the ``flipped" ${SU(5)\times U(1)}$ and $SU(6)\times SU(2)$ gauge groups in which gauge mediated proton decay operators are suppressed at leading order due to the special placement of matter fields in unified multiplets. We discuss both the theoretical structure and phenomenological implications of these models. For the latter, we examine the viability of the physical spectrum in each scenario and focus on the possible presence of other operators that could also contribute significantly to the proton decay rate.

 \vfill

\vspace{2cm}
\begin{flushleft}

May 2014
\end{flushleft}
\hrule width 6.7cm \vskip.1mm{\small \small}
 \end{titlepage}

\section{Introduction} The common and well-established perception of the Standard Model (SM) as an effective low energy theory of particle interactions has inevitably determined the direction of theoretical research over the past decades. In this framework, a number of interesting proposals, of varying elegance or virtue, have been put forward, that, in agreement with its well-tested predictions, aim to overcome the deficiencies of the SM. Among these proposals Grand Unified Theories(GUTs) have been singled out as a promising framework with a number of specific implications. Moreover, due to their consistency with other interesting and fruitful ideas such as supersymmetry(SUSY) and string theory, GUTs continue to draw interest for new theoretical realisations.

Unification of the fundamental forces is, in principle, a very compelling idea. The GUT approach, based essentially on the  mathematical and conceptual principles of a general Yang-Mills  theory, can be regarded as a minimal, yet non-trivial extension of the SM in this direction. GUTs offer simple and elegant answers to many of the puzzling situations met in the SM, such as an explanation for the charge quantization of elementary particles and a prospect for a unified description of the strong and electroweak interactions. In addition, the unification of gauge couplings of the MSSM could be interpreted as indirect GUT evidence. Nevertheless, explicit realisations usually face serious problems mainly associated with the observed proton stability. In fact, proton decay is a common prediction in $SU(5)$ related models \cite{RPP,Dorsner:2004xx,Dorsner:2012nq} which can be traced to the presence of new heavy particles that are unavoidably present due to the larger symmetry. These particles, charged under color and weak hypercharge, mediate baryon and lepton number violating processes through  higher dimensional operators. Current experimental tests on proton stability, on the other hand, impose stringent constraints on the presence of these operators. As a result many interesting GUT and SUSY-GUT minimal models are now either ruled out or extremely disfavored \cite{RPP,Bajc,Dorsner:2004jj}. It should be noticed however that even if baryon and lepton number violating operators are present the result is not always catastrophic. As it has been shown some time ago in a general approach\cite{DH} and also pointed out recently\cite{BARR,Barr:2014nza}, gauge-mediated proton decay can be severely suppressed well beyond current experimental bounds if the SM fermions are arranged properly in the GUT representations. In fact in \cite{DH} this issue was thoroughly and systematically investigated for various GUT groups with respect to the  gauge mediated $D=6$ operators. Of course, other sources of proton decay, such as fermion and scalar mediated $D=5,6$ operators, can prove equally dangerous or even disastrous. In fact, the $D=5$ operators are the major proton decay problem of SUSY-GUT models. Nevertheless, the presence and the effects of these operators are model dependent and should be thoroughly analyzed when investigating specific GUT models.

In what follows we investigate three possible SUSY-GUT realisations based on the $SU(5)\times U(1)_X$ and $SU(6)\times SU(2)_{L,R}$ gauge groups \cite{Slansky:1981yr}. There are strong motivations for the choice of the specific gauge groups. All of them allow for suitable fermion representations consistent with a heavy suppression of $D=6$ gauge mediated operators. In addition, the above gauge groups are favoured by string theory considerations \cite{RT2} and as subgroups of $E_6$ they also offer the possibility of further unification. We study these models from the viewpoint of proton decay and phenomenological viability of their spectrum. Since these are explicit realisations, contrary to previous more general approaches, technical difficulties associated with the theoretical structure unavoidably arise. However, it is interesting to notice that in some cases these problems can be evaded in a rather elegant manner.

\section{An Extended Flipped $SU(5)$ Model}
In this section we consider an extended supersymmetric $SU(5)\times U(1)_X$ model that implements the above ideas on proton decay suppression. The field content is that of the minimal version\cite{Barr:1981qv,Derendinger:1983aj,RT} extended with a pair of pentaplets\cite{BARR}, namely
\begin{equation}
\begin{array}{l}
{\cal{F}}_{(10,1)}\,=\,(q,\nu^c,D'^{c})\\
\,\\
\overline{f}_{(\overline{5},-3)}\,=\,(L',u^c)\\
\,\\
\ell^c_{(1,5)}\,=\,e^c\\
\,\\
{\cal{G}}_{(5,-2)}\,=\,(L,\overline{D'}^c)\\
\,\\
\overline{\cal{G}}_{(\overline{5},2)}\,=\,(\overline{L}',{D}^c)
\end{array}\,\,\,\,\,\,\,\,\,\begin{array}{l}
{\cal{H}}_{(10,1)}\,=\,(Q_H,N_H^c,D_H^c)\\
\,\\
\overline{\cal{H}}_{(\overline{10},-1)}\,=\,(\overline{Q}_H,\overline{N}_H^c,\overline{D}_H^c)\\
\,\\
h_{(5,-2)}\,=\,(h_d,\overline{\delta}_h^c)\\
\,\\
\overline{h}_{(\overline{5},2)}\,=\,(h_u,\delta_h^c)\,.
\end{array}\label{fmc}
\end{equation}
The assignment above includes additional primed fields that  will eventually become superheavy. This will turn out to be crucial for the suppression of the dangerous $D=6$ operators.

The GUT breaking $SU(5)\times U(1)_X\,\rightarrow\,SU(3)_C\times SU(2)_L\times U(1)_Y$ proceeds in the standard way through the vevs 
$$\langle N_H^c\rangle\,=\,\langle\overline{N}_H^c\rangle\,\neq\,0\,,\,$$
along the  SM singlet direction of the Higgs fields ${\cal{H}}_{(10,1)},\overline{\cal{H}}_{(\overline{10},-1)}$.
The remnants of the Higgs mechanism comprise a pair of triplets  $D_H^c,\,\overline{D}_H^c$ and one singlet $(N_H^c+\overline{N}_H^c)/\sqrt{2}$.

We consider the renormalisable superpotential of the form
\begin{equation}
 {\cal{W}}_0\,=\,{\cal{Y}}_u\,{\cal{F}}\overline{f}\,\overline{h}\,+\,{\cal{Y}}_D\,{\cal{F}}{\cal{G}}{\cal{H}}\,+\,{\cal{Y}}_L\,\overline{f}\,\overline{\cal{G}}\,{\cal{H}}\,+\,\mu\,{\cal{G}}\,\overline{\cal{G}}\,+\,\lambda\,{\cal{H}}{\cal{H}}h\,+\,\overline{\lambda}\,\overline{\cal{H}}\,\overline{\cal{H}}\,\overline{h}\,
 \label{spa}
\end{equation}
together with the following  non-renormalisable terms
\footnote{The non-renormalisable terms can arise from interactions with heavy states of a fundamental underlying theory, e.g. string theory. For example the first term can be derived from a renormalisable superpotential of the form ${\cal{F}}\overline{\cal{H}}\,S+\overline{\cal{G}}\,h\,S\,+\,M\,S^2$ after the integration of  the heavy singlet field $S$.}
\begin{equation}
\,\delta{\cal{W}}\,=\,\frac{{\cal{Y}}_d'}{M}{\cal{F}}\,\overline{\cal{G}}h\overline{\cal{H}}\,+\,\frac{{\cal{Y}}_e'}{M}{\cal{G}}\ell^ch\overline{\cal{H}}\,,
\label{spb}
\end{equation}
where we have suppressed family indices.

 The couplings $\lambda{\cal{H}}^2h+\overline{\lambda}\,\overline{\cal{H}}^2\overline{h}$ provide the mass terms
\begin{equation}\lambda\langle N_H^c\rangle\,D_H^c\,\overline{\delta}_h^c\,+\,\overline{\lambda}\,\langle\overline{N}_H^c\rangle\overline{D}_H^c\,\delta_h^c\label{flippedhiggs}
\end{equation}
that remove the color triplets $\delta_h^c,\,\overline{\delta}_h^c$ from the light spectrum, while the weak isodoublets $h_u,\,h_d$ remain massless.

Next, we focus on the superpotential couplings
\begin{eqnarray}
\mu\,{\cal{G}}\overline{\cal{G}}+{\cal{Y}}_D\,{\cal{F}}{\cal{G}}{\cal{H}}+{\cal{Y}}_L\,\overline{f}\,\overline{\cal{G}}{\cal{H}}&\supset &
\mu\left(L\overline{L'}+\overline{D'}^c{D}^c\right)+{\cal{Y}}_D\,\langle N_H^c\rangle D'^c\overline{D'}^c\,+\,{\cal{Y}}_L\langle N_H^c\rangle L'\overline{L'}\nonumber\\
&&=
\mu_L\,\overline{L'}\,{\cal{L}}\,+\,\mu_D\,\overline{D'}^c\,{\cal{D}}^c\,,{\label{MASS}}
\end{eqnarray}
and express them in terms of the mass eigenstates 
\begin{equation}
\begin{array}{cc}
{\cal{L}}\,\equiv\,\cos\theta_L\,L'\,+\,\sin\theta_L\,L\,&\,{\cal{D}}^c\,\equiv\,\cos\theta_D\,D'^c\,+\,\sin\theta_D\,{D}^c\,\\
\ell\,\equiv\,-\sin\theta_L\,L'\,+\cos\theta_L\,L&\,d^c\,\equiv\,-\sin\theta_D\,D'^c\,+\cos\theta_D\,{D}^c\end{array}
\end{equation}
where
\begin{equation}
\tan\theta_L\,\equiv\,{\cal{Y}}_L^{-1}\frac{\mu}{\langle N_H^c\rangle},\,\,\,\,\tan\theta_D\,\equiv\,{\cal{Y}}_D^{-1}\frac{\mu}{\langle N_H^c\rangle}\,.\label{tan}\end{equation}

Independently of the exact values of $\theta_L$ and $\theta_D$,
the expression ({\ref{MASS}}) suggests that the pairs ${\cal{L}},\,\overline{L'}$ and ${\cal{D}}^c,\,\overline{D'}^c$ acquire heavy masses $\mu_{L(D)}^2=\mu^2+{\cal{Y}}_{e(d)}^2\langle N_H^c\rangle^2$ while the states $\ell$ and $d^c$ are massless at the GUT level. 
The latter, however, will obtain  electroweak masses from the non-renormalisable couplings in $\delta{\cal{W}}$. As a result we have the following couplings with SM Higgs doublets
\begin{eqnarray}
&{\cal{Y}}_u{\cal{F}}\overline{f}\,\overline{h}\,+\,\frac{{\cal{Y}}_d'}{M}{\cal{F}}\overline{\cal{G}}h\overline{\cal{H}}\,+\,\frac{{\cal{Y}}_e'}{M}{\cal{G}}\ell^ch\overline{\cal{H}}&\supset{\cal{Y}}_u\left(\,q \,u^c\,+\,\nu^c L'\,\right)h_u\,+\,{\cal{Y}}_d'\frac{\langle\overline{N}_H^c\rangle}{M}q\,{D}^c\,h_d\,\nonumber\\
&&+\,{\cal{Y}}_e'\frac{\langle\overline{N}_H^c\rangle}{M}Le^ch_d
\end{eqnarray}
or in terms of the mass eigenstates
\begin{eqnarray}
&{\cal{Y}}_u\left(\,q \,u^c\,-\,\sin\theta_L\,\nu^c\ell\,+\,\cos\theta_L\,\nu^c \cal{L}\right)h_u\,
+{\cal{Y}}_d'\frac{\langle\overline{N}_H^c\rangle}{M}\left(\cos\theta_D \,q\,d^c + \sin\theta_D \,q\,{\cal{D}}^c \right)\,h_d&\,\nonumber\\
\,&+\,{\cal{Y}}_e'\frac{\langle\overline{N}_H^c\rangle}{M}\,\left(\cos\theta_L\,\ell\,e^c\,+\,\sin \theta_L \,{\cal{L}}\,e^c\right)h_d\,.&
\end{eqnarray}
Neglecting the couplings involving the heavy fields ${\cal{L}},\,{\cal{D}}^c\,$ we obtain the following
light fermion mass terms
\begin{equation}
{\cal{Y}}_u\left(\,q \,u^c\,-\,\sin\theta_L\,\nu^c\ell\,\right)h_u\,
+{\cal{Y}}_d'\frac{\langle\overline{N}_H^c\rangle}{M}\left(\cos\theta_D \,q\,d^c \,\right)\,h_d \,+ \,{\cal{Y}}_e'\frac{\langle\overline{N}_H^c\rangle}{M}\,\left(\cos\theta_L\,\ell\,e^c\,\right)h_d\,. \label{lfmass}
\end{equation}
In addition, a hierarchy between up quarks and the other charged fermions may be generated from the factor $\langle\overline{N}_H^c\rangle/{M}$, thus allowing for  small $\tan\beta$ values.

As far as proton decay is concerned, the only relevant gauge mediated $D=6$ operator  in this model is ${\cal{F}}{\cal{F}}^{\dagger}\overline{f}\,\overline{f}^{\dagger}$, giving for the light states
\begin{align}
q{D'^c}^{\dagger}L'{u^c}^{\dagger}\,\rightarrow\,\sin\theta_L\,\sin\theta_D\,\,q\,{d^c}^{\dagger}\ell\,{u^c}^{\dagger}\,.
\label{pto}
\end{align}
All other possible operators should be considered safe since they  involve at least one of the superheavy fields  $\overline{L}',\overline{D'}^c$. This is true for loop effects as well, since these are always followed by extra $(4\pi)^{-4}$ suppression factors.

It should be  clear  that in the limit
\begin{eqnarray}
\mu\,\ll\,{\cal{Y}}_L\langle N_H^c\rangle\,\longrightarrow\gap \theta_L\rightarrow\,0\nonumber\\
\mu\,\ll\,{\cal{Y}}_D\langle N_H^c\rangle\,\longrightarrow \gap\theta_D\rightarrow\,0\label{limit}
\end{eqnarray}
 the dangerous operator in \eqref{pto} is severely suppressed. 

The fact that we can so easily suppress proton decay in the limit \eqref{limit} should come as no surprise. This choice for the parameters corresponds to
\begin{eqnarray}
&{\cal L}\simeq \,L'\,,\,\,{\cal D}^c\,\simeq D'^c&\nonumber\\
&{\ell}\simeq \,L\,,\,\,{d}^c\,\simeq D^c\,,&
\end{eqnarray}
and as a result the light mass eigenstates $\ell,d^c,u^c$ and $q$ reside in different  $SU(5)$ representations. Therefore, each $SU(5)$ representation includes no more than one light field and a transition through the exchange of gauge bosons inevitably involves one heavy fermion state\footnote{For a more precise description of the $D=6$ suppression conditions see the criteria in \cite{DH}.}.

In the considered limit gauge mediated $D=6$ operators are safe, but there can be other sources of proton decay which could prove more dangerous.
These are  the  $D=5$ and the scalar mediated $D=6$ operators. The relevant terms from the superpotential read
\begin{eqnarray}
\Yu_u \Fu\overline{f}\,\overline{h}\,+\,\Yu_D \Fu\Gu\Hu \,+\,\Yu_L \overline{f}\overline{\Gu}\Hu &\supset & {\cal{Y}}_u\left(\,D'^cu^c\delta_h^c\,+\,qL'\delta_h^c\right)\,+\,{\cal{Y}}_D\left(\,q\,L\,D_H^c\,\right)\,
\nonumber\\&&\, {\cal{Y}}_L\left(\,u^c\,D^c\,D_H^c\,\right)\,
\end{eqnarray}
Clearly, from these terms no dangerous $D=5$ operator can be formed and the possibly dangerous scalar mediated $D=6$ operators are
\begin{eqnarray}
D'^c u^c(qL')^\dag\rightarrow \sin\theta_D\sin\theta_L \,d^c u^c(q L)^\dag \nonumber\\
qL (u^c D^c)^\dag\rightarrow \cos\theta_D\cos\theta_L \, q\ell (u^c d^c)^\dag\,. \label{flipd6}
\end{eqnarray}
The first is heavily suppressed for $\theta_L,\theta_D\rightarrow 0$ but the second will maximize in this limit. Nevertheless, this operator depends on both $\Yu_D,\Yu_L$ couplings which are relevant only for heavy matter and thus can be easily taken sufficiently small. The non-renormalisable part of the superpotential $\delta \Wu$ is irrelevant since these terms  are either heavily suppressed by the large mass $M$ or involve heavy fields.

The standard Yukawa couplings, which  have not been  included in the superpotential
 \eqref{spa},\eqref{spb} may also have important contributions to proton decay in 
 the above limit. 
 These are
 \begin{eqnarray}
 {\cal{Y}}_d \,{\cal{F}}\,{\cal{F}}\,h\,+\,{\cal{Y}}_e \,{\cal{\ell}}^c{\overline{f}}\,\,h\,\supset\,{\cal{Y}}_d\,qq\overline{\delta}_h^c\,+\,{\cal{Y}}_e\,e^cu^c\overline{\delta}_h^c
 \end{eqnarray}
 and as a result the  $D=5,\,6$ operators $qqq\ell ,\,e^cu^cu^cd^c,(qq)^\dag e^c u^c$ appear which are controlled by $\Yu_D\Yu_d,\,\Yu_L\Yu_e,\,\Yu_d^*\Yu_e$ respectively.
 Moreover, their contributions to fermion masses, are negligible 
\begin{eqnarray}
{\cal{Y}}_d \,{\cal{F}}\,{\cal{F}}\,h\,&\supset &\,{\cal{Y}}_d\,q\,D'^c\,h_d \,=\,{\cal{Y}}_d\,q\,(\cos\theta_D \,{\cal D}^c+ \sin\theta_D\, d^c)\,h_d\\
{\cal{Y}}_e \,{\cal{\ell}}^c{\overline{f}}\,\,h\,&\supset & {\cal{Y}}_e \,e^c\,L'\,h_d\,=\,{\cal{Y}}_e \,e^c\,(\cos\theta_L \,{\cal L}'+ \sin\theta_L\, \ell)\,h_d
\end{eqnarray}
 Of course, proton decay  can be tolerated as long as the couplings ${\cal{Y}}_d,\,{\cal{Y}}_e$ are also small.
However, a more drastic and perhaps more attractive solution would be to set these couplings to zero with the help of a symmetry.
 This is possible in this model since down-quark and charged lepton masses originate from another sector.

To this end we consider the superpotential
\begin{eqnarray}
{\cal{W}}_{R}\,&=&\,{\cal{Y}}_u\,{\cal{F}}\overline{f}\,\overline{h}\,+\,{\cal{Y}}_D\,{\cal{F}}\,{\cal{G}}{\cal{H}}\,+\,{\cal{Y}}_L\,\overline{f}\,\overline{\cal{G}}{\cal{H}}\,+\,\lambda\,{\cal{H}}{\cal{H}}h\,+\,\overline{\lambda}\,\overline{\cal{H}}\,\overline{\cal{H}}\,\overline{h}\,
\nonumber\\
&&+\,\frac{{\cal{Y}}_d'}{M}{\cal{F}}\,\overline{\cal{G}}h\overline{\cal{H}}\,+\,\frac{{\cal{Y}}_e'}{M}{\cal{G}}\ell^ch\overline{\cal{H}}
\end{eqnarray}
invariant under the ${\cal{Z}}_4^{(R)}$ symmetry\cite{PT-RSYM}
$$\begin{array}{cccc}
{\cal{F}},\,{\cal{G}}\,\rightarrow\,3\,&\,\,h,\,\overline{h}\,\rightarrow\,2\,&\,\overline{f},\,\ell^c,\,\overline{\cal{G}}\,\rightarrow\,1\,&\,\,{\cal{H}},\,\overline{\cal{H}}\,\rightarrow\,0\ .
\end{array}$$
Due to this symmetry not only the standard Yukawa  terms ${\cal Y}_d  {\cal F F}h,\,{\cal Y}_e \ell^c\overline{f} h$ are absent from the superpotential but also the mass term $\mu \,{\cal G}\overline{\cal G}$. The absence of the former results in the vanishing of the potentially dangerous operators, discussed previously. The absence of the latter however results in
\begin{equation}
\theta_L\,=\,\theta_D\,=\,0
\end{equation}
and the limit (\ref{limit}) is then automatically satisfied without any  assumption on the parameters.

There is also an option to include neutrinos in our discussion. However, that would unnecessarily restrict the model, offering, in most cases, no practical implication on the proton decay issue and furthermore no definite prediction on neutrino masses or mixing. In any case, for consistency of the spectrum, we may consider a rather minimal extension of the model above, by introducing a total singlet $N(1,0)$ with charge
$$N \,\rightarrow \, 1 $$
That would allow for the terms in the superpotential
\begin{equation}
\Wu_N\,=\, \Yu_\nu\, \Gu N\overline{h}\, +\, M_N N N
\end{equation}
and at the same time forbid $\overline{\Gu} Nh,\, \Gu\overline{\Gu}N,\,N\Hu\overline{\Hu},Nh\overline{h},\,N\Fu\overline{H}$.
This results in a \textit{type-I} seesaw with the typical light masses $m_\nu \simeq (\Yu_\nu v_u)^2/M_N$. Of course, the new operators arising will not affect the proton decay rate since they always involve heavy states. However, as previously mentioned, this is not the only viable extension for neutrinos and certainly not the most predictive one.

Summarizing, in this section we constructed and analysed an $SU(5)\times U(1)$ GUT 
with non-standard (\textit{deunified}) matter assignments. The usual lepton doublets and d-quark triplets
reside in pairs of additional $SU(5)$ vectorial multiplets, while
extra heavy matter fields are placed in their traditional locations in the chiral antisymmetric and vectorial $SU(5)$  representations. As a result, the gauge mediated $D=6$ Baryon and Lepton Number violating operators are suppressed. Moreover, proton decay through $D=5$ and scalar mediated $D=6$ 
operators can be evaded and protons are effectively stable. 
In addition,  we can obtain  a realistic light fermion mass spectrum.  We also discussed a minimal extension of the model that includes a viable neutrino spectrum without affecting the prediction for the  proton decay rate.

\section{An $SU(6)\times SU(2)_R$ Model}
Motivated by the attractive and elegant features of the extended flipped $SU(5)$ model discussed in the last section we investigate  possible embeddings of the MSSM  in larger symmetry groups. One 
possibility is the  $SU(6)\times SU(2)$ gauge group, a maximal subgroup of $E_6$, which also allows us to implement the ``deunification" scenario for proton decay suppression.
There are two possible embeddings of the weak isospin in this gauge group, namely $SU(6)\times {SU(2)}_L$ and  $SU(6)\times {SU(2)}_R$\cite{RT2}. However,
only the $SU(6)\times SU(2)_R$ gauge symmetry admits an $SU(5)\times {U(1)}_X $ subgroup. In this section we construct and analyse such a model where all flipped representations along with extra Higgs and matter fields are promoted into $SU(6)\times {SU(2)}_R$ multiplets. Despite the larger symmetry imposing stringent constraints on the parameters and the additional fields introduced, the light spectrum of the MSSM can be still obtained without any exotics.

We assume the following field content 
\begin{equation}
\begin{array}{l}
{\Psi}_{(15,1)}\,=\,({\cal F},{\cal G})\\
\,\\\,\\
{\psi}_{(\overline{6},2)}\,=\,(\ell^c,\overline{f},N,\overline{\cal G})
\end{array}\,\,\,\,\,\,\,\,\,\begin{array}{l}
{\Phi}_{(15,1)}\,=\,({\cal H},h_1)\\
\overline{\Phi}_{(\overline{15},1)}\,=\,(\overline{\cal H},\overline{h}_1)\\
\,\\
\phi_{(\overline{6},2)}\,=\,(\ell^c_H,\overline{f}_H,N_H,\overline{h}_2)\\
\overline{\phi}_{(6,2)}\,=\,(\overline{\ell}^c_H,f_H,\overline{N}_H,h_2)\,,
\end{array}
\end{equation}
where we employ the notation of \eqref{fmc}. As compared to the extended flipped model, there is an extra matter singlet field $\left(N\right)$  and a number of additional Higgs fields $\left(\ell^c_H,\overline{f}_H,N_H,\overline{\ell}^c_H,f_H,\overline{N}_H\right)$. GUT symmetry breaking is accomplished by the use of a two step Higgs mechanism, as follows
\begin{eqnarray}
SU(6)\times SU(2)_R\overset{\langle N_H,\overline{N}_H \rangle}{\longrightarrow} SU(5)\times U(1)_X \overset{\langle N^c_H,\overline{N}^c_H\rangle}{\longrightarrow} SU(3)_C\times SU(2)_L\times U(1)_Y\,,
\end{eqnarray}
 where $N^c_H,\overline{N}^c_H$ are the singlets residing in ${\cal{H}}_{(10,1)},\overline{\cal{H}}_{(\overline{10},-1)}$ of the 
 $SU(5)\times{U(1)}$ subgroup respectively and
\begin{eqnarray}
\langle N_H \rangle = \langle \overline{N}_H\rangle \geq \langle N_H^c \rangle=\langle \overline{N}_H^c\rangle\,.
\end{eqnarray}

The superpotential  
\begin{align}
W=\Wu_1+\Wu_2\ ,\label{su6r}
\end{align}
 where
\begin{eqnarray}
\Wu_1\,&=&\,{\cal Y}_D\,\,\Psi\Psi\Phi\,+\,{\cal Y}_L\,\psi\psi\Phi\,+\,\lambda_1 \Phibar^3\,+\,\lambda_2\,\phi^2\Phi\,\label{su6r1}\\
\Wu_2\,&=&\,\frac{\cal Y}{M}\Psi\psi\phibar\,\Phibar\,+\,\frac{\lambda'}{M}\Phi^2\phibar^2\,\label{su6nr}
\,,
\end{eqnarray}
is sufficient to guarantee the decoupling of the additional exotic states as well 
as  providing  mass terms for the SM fermions.
At the first step of symmetry breaking through a vev in the D-flat, flipped $SU(5)$ singlet direction, the fields $\ell^c_H,\overline{\ell}^c_H,f_H,\overline{f}_H,(N_H-\overline{N}_H)/\sqrt{2}\subset\phi ,\phibar$ are higgsed away leaving behind $h_2,\overline{h}_2$ and the orthogonal combination $(N_H+\overline{N}_H)/\sqrt{2}$. At this stage the relevant Higgs superpotential reduces to
\begin{equation}
\Wu_H\,=\,\lambda_1 \overline{\Hu}^2\overline{h}_1\,+\,\lambda_2\, \langle N_H \rangle\overline{h}_2h_1\,+\,\frac{\lambda'}{M}\langle\overline{N}_H\rangle \Hu^2 h_2
\,.
\end{equation}
It is clear that one pair of pentaplets ($h_1,\overline{h}_2$) becomes heavy. Hence, the light MSSM higgs doublets should reside in the other pair of pentaplets ($h_2,\overline{h}_1$) which remains massless at this level. For the second step of symmetry breaking down to the SM gauge group we may neglect the second term and work exactly as in the flipped $SU(5)$ case encountered in the previous section. The coloured triplets $\delta_{h_1}^c\in \overline{h}_1$, $\overline{\delta}_{h_2}^c\in h_2$  together with $\overline{D}^c_H, D^c_H$ acquire heavy masses of the order of 
$\lambda_1\,\langle N^c_H \rangle$ and $\lambda'\,\langle \overline{N}_H \rangle\langle N^c_H \rangle/M$ respectively 
leaving behind only the massless doublets $h_u$ and $h_d$.
Obviously there are no exotic remnants in the Higgs sector. 

We next focus on the matter sector of (\ref{su6r}). In terms of flipped $SU(5)$
multiplets we have the following decompositions of the relevant terms
\begin{eqnarray}
\Psi\Psi\Phi &=& {\cal F G H}+{\cal F}^2 h_1 \label{su6ryd}\\
\psi\psi\Phi &=& {\overline f}\overline{\Gu}\Hu + {\overline f}\ell^c h_1 + \overline{\Gu} N h_1 \\
\Psi\psi\phibar\,\Phibar &=& \Gu\overline{\Gu}h_2\overline{h}_1+\Fu\overline{f}\overline{N}_H\overline{h}_1+\Gu\ell^ch_2\overline{\Hu}+\Fu\overline{\Gu} h_2\overline{\Hu}\nonumber\\ &&+\Gu N\overline{N}_H\overline{h}_1 + \Fu N \overline{N}_H\overline{\Hu}\,.\label{su6ry}
\end{eqnarray}
The light charged fermion masses arise from the couplings\footnote{Charged lepton and down quark masses also receive negligible contributions from $\frac{\cal Y}{M}\Gu\overline{\Gu}h_2\overline{h}_1\,$.}
\begin{eqnarray}
\textrm{up quarks  } &:&   \frac{\Yu}{M} \Fu\overline{f}\overline{N}_H\overline{h}_1
\sim \Yu\,\frac{\langle \overline{N}_H\rangle}{M} \Fu\overline{f}\,\overline{h}_1
\\
\textrm{down quarks  } &:&  \frac{\Yu}{M}\Fu\overline{\Gu} h_2\overline{\Hu}
\sim \Yu\,\frac{\langle\overline{N}_H^c\rangle}{M}\Fu\overline{\Gu} h_2
\\
\textrm{charged leptons  }&:&  \frac{\Yu}{M}\Gu\,\ell^c h_2\overline{\Hu}
\sim \Yu\,\frac{\langle\overline{N}_H^c\rangle}{M}\Gu\ell^c h_2
\, .
\end{eqnarray}
Since $h_1,\overline{h}_2$  decouple at GUT scale, operators involving them are irrelevant for fermion masses. 

The terms associated with proton suppression mechanism  are
\begin{eqnarray}
\Wu_M\,=\,{\cal Y}_D\,\,{\cal F G H}+\,{\cal Y}_L\,{\overline f}\,\overline{\Gu}\Hu +\,\frac{\cal Y}{M}\Gu\overline{\Gu}h_2\overline{h}_1\,.
\end{eqnarray}
This is essentially expression (\ref{MASS}) with $\mu\equiv {\cal Y}\frac{v_u v_d}{M}$. However, here
\begin{equation}
{\cal Y}\frac{v_u v_d}{M}\ll \Yu_L\langle N_H\rangle \sim \Yu_D \langle N_H\rangle
\,,
\end{equation}
satisfies automatically the  condition \eqref{limit} , i.e. $\theta_L,\,\theta_D\rightarrow\, 0$. As in the extended flipped $SU(5)$ case,
this guarantees both the decoupling of extra matter and the suppression of
dangerous gauge mediated $D=6$ operators. Hence, the ``de-unification" scenario 
can be also realised in the $SU(6)\times SU(2)_R$ case despite the additional 
gauge symmetry. This can be seen as follows: The standard matter is distributed as
\begin{eqnarray}
q,\nu^c,\ell\in \Psi &\gap\gap \gap & e^c,u^c\in \psi
\end{eqnarray}
As a result, the gauge mediated $D=6$ baryon or lepton number violating operators will necessarily form, at tree level, as products  of the bilinears
\begin{eqnarray}
(q\,\ell^\dag)\gap,\gap(q^\dag\ell)\gap,\gap({u^c}^\dag e^c)\gap ,\gap(u^c {e^c}^\dag)\ .
\end{eqnarray}
However, gauge symmetry forbids the appearance of all relevant combinations.

Next, we investigate the presence of other dangerous baryon decay operators. As seen from (\ref{su6ryd}-\ref{su6ry}) the relevant terms are
\begin{eqnarray}
\,{\cal Y}_D\,\,\Psi\Psi\Phi\,+\,{\cal Y}_L\,\psi\psi\Phi\,+\,\frac{\cal Y}{M}\Psi\psi\phibar\,\Phibar &\supset &\Yu_D (\,qLD^c_H+qq\,\overline{\delta}^c_{h_1} )\,+\,\Yu_L(\,u^c D^c D^c_H\,+\,u^ce^c\overline{\delta}^c_{h_1})\nonumber\\
&&+\,\Yu\frac{\langle\overline{N}_H\rangle}{M}\left(\, D'^cu^c\delta_{h_1}^c\,+\,qL'\delta_{h_1}^c\,\right)
\end{eqnarray}
Since in this model there are no  $\overline{\delta}^c_{h_1} {\delta}^c_{h_1}$ or $\overline{\delta}^c_{h_1} D^c_{H}$ mass terms, the potentially dangerous operators are only the $D=6$  of (\ref{flipd6}) and the new $qq(u^c e^c)^\dag$ mediated by $\overline{\delta}^c_{h_1}$. These operators can be easily suppressed taking $\Yu_D\Yu_L\ll 1$, exactly as in the flipped case, since the Yukawa couplings $\Yu_D\,,\Yu_L$ are not related to 
charged fermion masses.

\section{ $ SU(6)\times SU(2)_L$ Models }
Our next, rather obvious step, is to investigate the above ideas for proton decay suppression  in the alternative distinct embedding of the SM gauge group namely in $ SU(6)\times SU(2)_L$. This embedding is not only interesting for reasons of completeness  but also for its unique matter multiplet structure. In this scenario, standard matter is a priori \textit{``deunified''} as
$$e^c,u^c,d^c \in \Psi_{\left(\overline{{15}},{1}\right)}\ ,\ q,l\in \psi _{\left({{6}},{2}\right)}\,.$$
  Clearly this is a good starting point for model building since no dangerous, gauge mediated, $D=6$ operator can
 be formed from the bilinears
\begin{eqnarray}
(q\,\ell^\dag)\gap,\gap(q^\dag\ell)\gap,\gap({u^c}^\dag e^c)\gap ,\gap(u^c {e^c}^\dag)\gap , \gap({u^c}^\dag d^c)\gap ,\gap(u^c {d^c}^\dag)\gap , \gap({e^c}^\dag d^c)\gap ,\gap(e^c {d^c}^\dag)
\end{eqnarray}
at tree level. 
Furthermore, there are two possible models depending on the assignment of the light Higgs doublets ($h_u,h_d$). They may either reside in a pair of additional 
$\left({\mb{6}},\mb{2}\right)+\left({\overline{\mb{6}}},\mb{2}\right)$ multiplets (Model I) or 
together with $q,\ell$ in chiral  $\left({{\mb{6}}},\mb{2}\right)$ multiplets (Model II). In the former case we have  two ($h_u,h_d$) pairs while in the latter we have three ($h_u,h_d$) pairs.

In what follows we investigate both models, each with different  predictions for the spectrum and proton decay.\newline 

{\noindent \bf Symmetry Breaking\newline}
For both models discussed in this section, we consider the GUT breaking chain
\begin{equation}
SU(6)\times SU(2)_L\,\overset{\langle N_1,\overline{N}_1\rangle}{\rightarrow}\,SU(4)\times SU(2)_R\times SU(2)_L\,\overset{\langle\overline{N}^c_2,N^c_2\rangle}{\rightarrow}\,SU(3)_C\times U(1)_Y\times SU(2)_L\,.\nonumber
\end{equation}
This can be accomplished with the help of three pairs of GUT-Higgs multiplets denoted as $\Phi_i,\,\overline{\Phi}_i$ and transforming as
\begin{eqnarray}
{\Phi}_i(\overline{15},1)\,=\,(N_i)_{(1,1,1)}\,+\,\left(\Delta_i\,+\,\Delta_i^c\right)_{(6,1,1)}\,+\,\left(E_i^c+N_i^c+D_i^c+U_i^c\right)_{(\overline{4},2,1)}\nonumber\\
\overline{\Phi}_i(15,1)\,=\,(\overline{N}_i)_{(1,1,1)}\,+\,\left(\overline{\Delta}_i\,+\,\overline{\Delta}_i^c\right)_{(6,1,1)}\,+\,\left(\overline{E}_i^c+\overline{N}_i^c+\overline{D}_i^c+\overline{U}_i^c\right)_{({4},2,1)}
\end{eqnarray}
in terms of $SU(6)\times SU(2)_L$ (left) and $SU(4)\times SU(2)_R\times SU(2)_L$ (right)\footnote{We also use an implicit $SU(3)\times SU(2) \times U(1)$ field notation. In this notation, the extra fields introduced here will transform as $\Delta^c,\overline{\Delta} \sim D^c(\overline{3},1,1/3)$ and $\overline{\Delta}^c,{\Delta} \sim \overline{D}^c(3,1,-1/3)$.}.

We assume the following GUT-Higgs superpotential
\begin{equation}
{\cal{W}}_\Phi\,=\,\lambda_{122}\Phi_1{\Phi}_2^2\,+\,\lambda_{113}{\Phi}_1^2{\Phi}_3\,+\,\overline{\lambda}_{122}\overline{\Phi}_1\,\overline{\Phi}_2^2\,+\,
\overline{\lambda}_{113}\overline{\Phi}_1^2\,\overline{\Phi}_3\,+\,M{\Phi}_3\overline{\Phi}_3\,.\label{su6lWPhi}
\end{equation}
The first step of symmetry breaking is realised through a vev in the F-,D- flat direction
\begin{equation}
\langle N_1\rangle\,=\,\langle\overline{N}_1\rangle\,=\,V_1\,\neq\,0
\end{equation}
All components of $\Phi_1,\,\overline{\Phi}_1$ are then higgsed away except $(N_1+\overline{N}_1)/\sqrt{2},\,\Delta_1,\,\Delta_1^c,\,\overline{\Delta}_1,\,\overline{\Delta}_1^c$. The second step of symmetry breaking down to the SM is realised through a vev in the direction
\begin{equation}
\langle N_2^c\rangle\,=\,\langle \overline{N}_2^c\rangle\,=\,V_2\,\neq\,0,
\end{equation}
leaving as remnants in $\Phi_2,\, \Phibar_2$ the fields $(N_2^c+\overline{N}_2^c)/\sqrt{2},\,N_2,\,\overline{N}_2,\,\Delta_2,\,\Delta_2^c,\,\overline{\Delta}_2,\,\overline{\Delta}_2^c,\,D_2^c,\,\overline{D}_2^c$. The extra GUT-Higgs pair $\Phi_3 , \,\Phibar_3$ will not acquire a non vanishing vev in any direction, thus preserving the F-flatness of the superpotential. However its presence is required to render all Higgs remnants massive. The relative coupling obtained from the Higgs superpotential in (\ref{su6lWPhi})
\begin{eqnarray}
&&\lambda_{122}V_1\,\Delta_2\Delta_2^c\,+\,\lambda_{122}V_2\,D_2^c\Delta_1\,+\,\lambda_{113}V_1\left(\Delta_1\Delta_3^c+\Delta_3\Delta_1^c\right)\,\nonumber\\
&&+\,\overline{\lambda}_{122}V_1\,\overline{\Delta}_2\overline{\Delta}_2^c\,+\,\overline{\lambda}_{122}V_2\,\overline{D}_2^c\overline{\Delta}_1\,+\,\overline{\lambda}_{113}V_1\left(\overline{\Delta}_1\overline{\Delta}_3^c+\overline{\Delta}_3\overline{\Delta}_1^c\right)\,\nonumber\\
&&+M\left(\Delta_3\overline{\Delta}_3+\Delta_3^c\overline{\Delta}_3^c+D_3^c\overline{D}_3^c+U_3^c\overline{U}_3^c+\dots\right)
\end{eqnarray}
provide masses to all non-singlet GUT-Higgs field remnants.

{\noindent \bf Model- I\newline}
First, we consider the model where the  Higgs doublets belong to separate representations from standard matter. We thus introduce the Higgs pair
\begin{eqnarray}
\phi_{\left({6},{2}\right)} & = & (h+h^c)_{(1,2,2)} \, + \,(Q_H + L_H)_{(4,1,2)} \nonumber\\
\phibar_{(\overline{{6}},{2})} & = & (\overline{h}+\overline{h}^c)_{(1,2,2)} \, + \,(\overline{Q}_H + \overline{L}_H)_{(\overline{4},1,2)}\,,
\end{eqnarray}
while standard matter resides in
\begin{eqnarray}
\Psi_{\left(\overline{{15}},{1}\right)}&=&\nu_{(1,1,1)}\,+\,\left(\delta\,+\,\delta^c\right)_{(6,1,1)}\,+\,\left(\nu^c+e^c+d^c+u^c\right)_{(\overline{4},2,1)}\nonumber\\
\psi_{\left({6},{2}\right)}&=&\left(\eta\,+\,\eta^c\right)_{(1,2,2)}\,+\,\left(\ell+q\right)_{(4,1,2)}\,.
\end{eqnarray}
For reasons that will become apparent later we introduce  additional  matter in the self-conjugate $X\left(\mb{20},\mb{2}\right)$ representation.

We consider the superpotential relevant to the matter spectrum is
\begin{eqnarray}
\Wu_M &=& \Yu \Psi\psi\phi\,+\,\Yu' \Psi\Psi \Phi_1 \,+\,\Yu''\psi\psi\Phi_1\nonumber\\
&&+M_X X^2+\lambda X\Phibar_1\phi +\overline{\lambda} X\Phi_1\phibar +\lambda'\Phibar_1\phibar^2\label{su6lwm}
\end{eqnarray}
Here we restrict to one family description as the generalization to the three family case is straight-forward and can be obtained by taking copies of the standard matter multiplets.
The terms in the first row of \eqref{su6lwm} are responsible for the light matter masses 
as well as for decoupling of the extra non-MSSM states. In particular, we have
\begin{eqnarray}
\Yu\, \Psi\psi \phi &\supset & \Yu (\,\ell \, \nu^c h^c\,+\,\ell \, e^c h\,+\,q\,d^ch\,+\,q\, u^ch^c\,+\,\nu\eta h^c\,+\,\nu\eta^c h\,)\label{su6ly} \\
\Yu'\, \Psi\Psi\Phi_1 &\supset & \Yu' \delta \delta^c \langle N_1 \rangle  \\
\Yu''\, \psi\psi\Phi_1 &\supset & \Yu'' \eta \eta^c \langle N_1 \rangle
\end{eqnarray}
where if the light higgs doublets are identified as $h\equiv h_d,\, h^c\equiv h_u$ then (\ref{su6ly}) will include the standard Yukawa couplings of light matter. Furthermore, due to the $\Yu',\,\Yu''$ terms extra non-MSSM matter decouples leaving behind only singlets.

Terms in the second row of (\ref{su6lwm}) will induce the decoupling of extra $Q_H,L_H,\overline{Q}_H,\overline{L}_H$ in $\phi , \,\phibar$. Assuming $M_X\sim M_{pl}$, $X$ decouples leaving behind the effective mass operator
\begin{equation}\frac{\lambda\lambda'}{M_X}N_1\overline{N}_1\left(Q_H\overline{Q}_H\,+\,L_H\overline{L}_H\right)\,\Longrightarrow\,\frac{\lambda\lambda'\langle N_1\rangle^2}{M_X}\left(Q_H\overline{Q}_H\,+\,L_H\overline{L}_H\right)
\end{equation}
Hence, the $X$ fields are required in order to generate mass terms for the additional $Q_H,L_H$ type fields while keeping the associated Higgs doublets massless.

In addition the extra doublets in $\phibar$ will become superheavy through the couplings
\begin{equation}
\lambda'\Phibar_1\phibar^2\rightarrow \lambda'\langle N_1 \rangle\overline{h}\,\overline{h}^c
\end{equation}
and thus $h,h^c\in \phi$ are identified as the light Higgs doublets of the MSSM.

Altogether, there are no exotic remnants in this model and gauge mediated proton decay $D=6$ operators are  absent. However, there are other sources of proton decay. The potentially dangerous terms involving light matter are
\begin{eqnarray}
\Yu'\, \Psi\Psi\Phi_1 &\supset & \Yu' (\,e^cu^c\Delta_1\,+\,d^c u^c\Delta_1^c\,) \\
\Yu''\, \psi\psi\Phi_1 &\supset & \Yu'' (\,qq\Delta_1\,+\,q\ell\Delta^c_1 \,)\,.\label{su6ly''}
\end{eqnarray}
Since there is no $\Delta_1^c\Delta_1$ mass term, the associated $D=5$ operators will be also absent. On the other hand, the scalar mediated $D=6$ operators $q^\dag q^\dag e^cu^c,\,q^\dag \ell^\dag d^c u^c$ will be present but as in all previous models they can be suppressed by appropriate choice of $\Yu',\,\Yu''$. This is possible here also, since these couplings are only relevant to the decoupling scale of heavy matter.

{\noindent \bf Model- II\newline}
An alternative model can be obtained by assigning the higgs doublets of the MSSM to $\psi \left(\mb{6},\mb{2}\right)$ of matter. The spectrum in this case, besides the GUT Higgs fields $\Phi_i,\Phibar_i$, includes only the multiplets
\begin{eqnarray}
{\Psi_i}_{(\overline{{15}},{1})}&=&{\nu_i}_{(1,1,1)}\,+\,\left(\delta_i\,+\,\delta^c_i\right)_{(6,1,1)}\,+\,\left(\nu^c_i+e^c_i+d^c_i+u^c_i\right)_{(\overline{4},2,1)}\nonumber\\
{\psi_i}_{({6},{2})}&=&(h^u_i\,+\,h^d_i)_{(1,2,2)}\,+\,\left(\ell_i +q_i\right)_{(4,1,2)}
\end{eqnarray}
No additional matter fields are required in this case, which is certainly an improvement with respect to Model I.

The matter part of the superpotential is
\begin{equation}
\Wu_M = \Yu_{ijk} \psi_i\psi_j\Psi_k\,+\,\Yu'_{ij}\Psi_i\Psi_j\Phi_1\,+\, \Yu_{ijk}''\psi_i\psi_j\Phi_1
\end{equation}
out of which  the mass terms for light and heavy matter can be derived from\footnote{For simplicity, we neglect irrelevant numerical factors.}
\begin{eqnarray}
{\cal{Y}}_{ijk}\,\psi_i\psi_j\Psi_k &\supset&{\cal{Y}}_{ijk}\left(h_i^u\ell_j \nu_k^c \,+\, h^d_i\ell_je_k^c\,+\,h^d_iq_jd_k^c\,+\,h_i^uq_ju_k^c\right)\label{su6lyukawa}\\
\Yu'_{ij}\Psi_i\Psi_j\Phi_1 &\supset& {\cal{Y}}_{ij}'\,\delta_i\delta_j^c\langle N_1\rangle\label{su6ldelta}\\
\Yu''_{ij}\psi_i\psi_j\Phi_1 &\supset& {\cal{Y}}_{ij}''\,h^u_ih_j^d\langle N_1\rangle\label{su6lhiggs}
\end{eqnarray}
In order to obtain the MMSM spectrum at low energies we have to resolve the problem of the SM higgs  doublets introduced along with fermion generations. 
The simplest way to
achieve this is to make assumptions on the structure of the  coupling  $\Yu_{ij}''$.
Actually it is sufficient to assume that  $\Yu_{ij}''$ is a rank-2 symmetric matrix. 
In this case two linear combinations of Higgs pairs doublets will become superheavy (of the order of $\langle N_1\rangle$) and decouple from the light spectrum while one remains light. 
On the other hand, the light fermion spectrum can be directly obtained from the $\Yu_{ijk}$ coupling with the light Higgs states. In the general case, that would involve a rotation of the couplings to the Higgs mass eigenstate basis through $h^{(u,d)}_i=U_{ia}{h'}^{(u,d)}_a$ that would  diagonalize $\Yu''$ and bring (\ref{su6lyukawa}) to the form
\begin{equation}
U_{ia}{\cal{Y}}_{ijk}\left({h'}_a^u\ell_j \nu_k^c \,+\, {h'}^d_a\ell_je_k^c\,+\,{h'}^d_aq_jd_k^c\,+\,{h'}_a^uq_ju_k^c\right)\,.
\label{su6lyuk}
\end{equation}
If we identify the massless Higgs pair for $a=1$ then the Yukawa couplings of light matter will be given by 
\begin{equation}
Y_{jk}\,\equiv\,U_{i1}{\cal{Y}}_{ijk}\,,\label{su6lyuk1}
\end{equation}
and \eqref{su6lyuk} provides the standard fermion mass terms.


The presence of dangerous non-gauge mediated  operators is determined by the couplings
\begin{eqnarray}
{\cal{Y}}_{ijk}\,\psi_i\psi_j\Psi_k &\supset&{\cal{Y}}_{ijk}\left( q_iq_j\delta_k\,+\,q_i\ell_j\delta^c_k\right)\label{su6ldeltak}\\
\Yu'_{ij}\, \Psi_i\Psi_j\Phi_1 &\supset & \Yu'_{ij} (\,e^c_iu^c_j\Delta_1\,+\,d^c_i u^c_j\Delta_1^c\,)\\
\Yu''_{ij}\, \psi_i\psi_j\Phi_1 &\supset & \Yu''_{ij} (\,q_iq_j\Delta_1\,+\,q_i \ell_j\Delta_1^c\,)\,.
\end{eqnarray}
Scalar mediated $D=6$ operators will  emerge exactly as in the previous model. Nevertheless,  they are controlled again by the couplings $\Yu',\Yu''$ , which are not related to light matter.
As a result they can be easily suppressed. On the other hand the situation for $D=5$ operators is different. Although, as previously mentioned, a dangerous effective operator cannot be formed through the mediation of $\Delta_1\Delta_1^c$ since the associated mass term is absent, in this alternative model there is a new source of proton decay.  The terms in (\ref{su6ldeltak}) can in principle induce an effective dangerous $qqq\ell$ term through the mediation of $\delta_i\delta^c_j$. Moreover, the effective coupling of this higher dimensional operator is related to the standard Yukawa couplings of light matter and thus cannot be taken arbitrarily small. However, there is still an escape due to the family structure of the $\delta_i,\delta^c_i$. To understand this we may focus on the terms 
\begin{eqnarray}
{\cal{Y}}_{ijk}\left( q_iq_j\delta_k\,+\,q_i\ell_j\delta^c_k\right)\,+\,\Yu'_{ij}\delta_i\delta^c_j \langle N_1 \rangle
\end{eqnarray}
where the mass matrix for the heavy triplets is identified as the symmetric 
$M_{ij} \equiv \Yu'_{ij}\langle N_1 \rangle$. 
Then, as has been shown in \cite{RGT} the coupling of the $qqql$ operator will be given by 
\begin{eqnarray}
{\cal{O}}_{ijkl}^{qqql}\,=
\,\Yu_{ijp}\frac{(\textrm{cof}M)_{\,pq}}{\det{M}}\,\Yu_{klq}\,,
\end{eqnarray}
where $\textrm{cof}\left(M\right)$ is the matrix of cofactors for $M$. For a  symmetric texture of the form 
$$M\sim\left(\begin{array}{ccc}
0&0&a\\
0&c&b\\
a&b&d\\
\end{array}\right)$$
the mass matrix M will have the properties
$$\textrm{cof}\left(M_{33}\right)=0\,,\,\,\det{M}\neq 0$$
The dangerous operator will then be absent for  $\Yu_{ij1}=\Yu_{ij2}=0\,,\,\,\Yu_{ij3}\neq 0$, a condition which also predicts  Yukawa unification for the third family and a massless spectrum for the other two as can be seen from (\ref{su6lyuk1}). If we desire the suppression of this operator instead of its absence we could replace the above condition with $\Yu_{ij1}\sim\Yu_{ij2}\ll\Yu_{ij3}$ which would render the first two families massive but lighter than the third. Nevertheless, as happens with models predicting Yukawa unification, the presence of additional mass corrections is required in order to obtain realistic spectrum. 

In summary, we demonstrated that the proton decay problem can be also resolved in this model under some  assumptions on the structure of Yukawa couplings.

\section{Conclusions}
Grand Unified Theories provide a natural framework for extending the MSSM gauge interactions to a unified theory. Their predictions include coupling unification, charge quantization and some successful fermion mass relations. Moreover, the MSSM matter particles fit nicely into some of the lowest gauge group representations, as the $\mb{10}+\overline{\mb{5}}$ of $SU(5)$ or the $\mb{16}$ of $SO(10)$. However, one of their main consequences, namely nucleon decay, is not confirmed so far by experiments. This raises 
the question whether we could trade some of the GUT advantages for proton longevity.

In this paper, we have focused on the matter ``de-unification" scenario which amounts to distributing 
the light MSSM matter over several GUT gauge group representations together with additional heavy particles that  decouple at low energies.
We have implemented this idea in the context of three concrete  models, namely $SU(5)\times U(1)_X,\,SU(6)\times SU(2)_R$ and  $SU(6)\times SU(2)_L$. We have demonstrated that this non-minimal 
matter assignment  leads to a suppression of all dangerous gauge mediated $D=6$ operators, typically present in $SU(5)$ related GUTs. Moreover, the models  discussed are free of proton decay inducing $D=5$ operators while  dangerous scalar mediated $D=6$ operators
are under control since the associated couplings are only related to heavy matter.

Despite the non-standard matter assignments and the extra fields introduced we have shown that the models under consideration are free of exotics at the limit of low energies and   yield realistic charged fermion mass couplings. Combined with proton stability these features are highly non-trivial as even in the standard flipped $SU(5)$ scenario the characteristic doublet-triplet mechanism is not a priori expected to realise, once one departs from the minimal case.
However here, a consistent spectrum is always obtained and these models seem to offer a realistic escape from standard GUT problems.

\section*{Acknowlegments}
This research has been co-financed by the European Union (European Social Fund - ESF) and Greek
national funds through the Operational Program `` Education and Lifelong Learning"  of the National Strategic
Reference Framework (NSRF) - Research Funding Program: THALIS Investing in the society of knowledge
through the European Social Fund. AEF would like to thank the University of Ioannina for hospitality and the STFC (grant ST/J000493/1) for support.

\end{document}